\documentclass[emulateapj,natbib]{emulateapj}

\begin{document}
\title{Recovery of the Candidate Protoplanet HD 100546 \lowercase{b} with Gemini/NICI and Detection of Additional (Planet-Induced?) Disk Structure at Small Separations}
\author{
Thayne Currie\altaffilmark{1},
Takayuki Muto\altaffilmark{2},
Tomoyuki Kudo\altaffilmark{1},
Mitsuhiko Honda\altaffilmark{3},
Timothy D. Brandt \altaffilmark{4},
Carol Grady \altaffilmark{5},
Misato Fukagawa\altaffilmark{6},
Adam Burrows\altaffilmark{7},
Markus Janson\altaffilmark{8},
Masayuki Kuzuhara\altaffilmark{9},
Michael W. McElwain \altaffilmark{10},
Katherine Follette \altaffilmark{11},
Jun Hashimoto\altaffilmark{12},
Thomas Henning\altaffilmark{13},
Ryo Kandori\altaffilmark{14},
Nobuhiko Kusakabe\altaffilmark{14},
Jungmi Kwon\altaffilmark{15},
Kyle Mede\altaffilmark{15},
Jun-ichi Morino\altaffilmark{14},
Jun Nishikawa\altaffilmark{14},
Tae-Soo Pyo\altaffilmark{1},
Gene Serabyn\altaffilmark{16},
Takuya Suenaga\altaffilmark{14},
Yasuhiro Takahashi\altaffilmark{14,15},
John Wisniewski\altaffilmark{12},
Motohide Tamura\altaffilmark{14,15}
}
{
\altaffiltext{1}{NAOJ, Subaru Telescope, 650 N' Aohoku Pl., Hilo, HI 96720, \email{currie@naoj.org}}
\altaffiltext{2}{Kogashin University}
\altaffiltext{3}{Kanagawa University}
\altaffiltext{4}{Astrophysics Department, Institute for Advanced Study, Princeton, NJ, USA}
\altaffiltext{5}{Eureka Scientific, 2452 Delmer, Suite 100, Oakland CA96002, USA}
\altaffiltext{6}{Graduate School of Science, Osaka University, 1-1 Machikaneyama, Toyonaka, Osaka 560-0043, Japan}
\altaffiltext{7}{Department of Astrophysical Sciences, Princeton University, 7 Ivy Lane, Princeton, NJ}
\altaffiltext{8}{Stockholm University}
\altaffiltext{9}{Department of Earth and Planetary Sciences, Tokyo Institute of Technology, 2-12-1 Ookayama, Meguro-ku,
Tokyo 152-8551, Japan}
\altaffiltext{10}{Exoplanets and Stellar Astrophysics Laboratory, Code 667, Goddard Space Flight Center, Greenbelt, MD
20771, USA}
\altaffiltext{11}{Department of Astronomy, Steward Observatory, University of Arizona}
\altaffiltext{12}{H. L. Dodge Department of Physics \& Astronomy, University of Oklahoma, 440 W Brooks St Norman, OK 73019, USA}
\altaffiltext{13}{Max Planck Institute for Astronomy, K¬onigstuhl 17, 69117 Heidelberg, Germany}
\altaffiltext{14}{National Astronomical Observatory of Japan, 2-21-1, Osawa, Mitaka, Tokyo, 181-8588, Japan}
\altaffiltext{15}{Department of Astronomy, The University of Tokyo, 7-3-1, Hongo, Bunkyo-ku, Tokyo, 113-0033, Japan}
\altaffiltext{16}{Jet Propulsion Laboratory, 4800 Oak Grove Drive, MS 183-900, Pasadena, CA 91109, USA}

}
\begin{abstract}
We report the first independent, second-epoch (re-)detection of a directly-imaged protoplanet candidate.  Using 
$L^\prime$ high-contrast imaging of HD 100546 taken with the Near-Infrared Coronagraph and Imager (NICI) on Gemini South, we recover `HD 100546 b' with a position and brightness consistent with the original VLT/NaCo detection from Quanz et al, although data obtained after 2013 will be required to decisively demonstrate common proper motion.   HD 100546 b may be spatially resolved, up to $\approx$ 12-13 AU in diameter, and is embedded in a finger of thermal IR bright, polarized emission extending inwards to at least 0\farcs{}3.  
Standard hot-start models imply a mass of $\approx$ 15 $M_{J}$.  But if HD 100546 b is newly formed or made visible by a circumplanetary disk, both of which are plausible, its mass is significantly lower (e.g. 1--7 $M_{J}$).
Additionally, we discover a thermal IR-bright disk feature, possibly a spiral density wave, at roughly the same angular separation as HD 100546 b but 90 degrees away.   Our interpretation of this feature as a spiral arm is not decisive,  but modeling analyses using spiral density wave theory implies a wave launching point exterior to $\approx$ 0\farcs{}45 embedded within the visible disk structure:  plausibly evidence for a second, hitherto unseen wide-separation planet.
With one confirmed protoplanet candidate and evidence for 1--2 others, HD 100546 is an important evolutionary precursor to intermediate-mass stars with multiple super-jovian planets at moderate/wide separations like HR 8799.
\end{abstract}
\section{Introduction}
Nearly all stars are born surrounded by gas and dust-rich \textit{protoplanetary disks} comprising the building blocks for young 
gas giant planets \citep[e.g.][]{Hernandez2007,Cloutier2014}.  Over the past 7 years, direct imaging observations have revealed a dozen fully-formed (candidate) super-jovian planets (5--15 $M_{J}$)
around young stars \citep[e.g.][]{Marois2008a,Marois2010a,Lagrange2010,Carson2013,Kuzuhara2013,Rameau2013,Currie2014a}.
Direct images of \textit{protoplanets} embedded in disks provide a critical link between planet formation's starting point and these various giant planet formation outcomes.  


Numerous protoplanetary disks show strong but indirect evidence for infant jovians
\citep[e.g.][]{Muto2012,Grady2013,Quanz2013a,Garufi2013}.  However, no protoplanets have yet been independently confirmed in multiple epochs.
Some candidates -- e.g. T Cha b, LkCa 15 b -- are identified from nonzero closure phase signals from sparse aperture masking observations \citep[][]{Huelamo2011,Kraus2012}, but have yet to be confirmed by other authors or can be misidentified disk signals \citep{Olofsson2013,Cieza2013}.  
Two data sets  \citep{Reggiani2014,Biller2014} identified the candidate HD 169142 b at $r$ $\approx$ 1--2 $\lambda$/D (but at conflicting positions) with the same setup (camera, $t_{int}$, processing) and at the same epoch.   While promising,  this candidate is therefore susceptible to the same misinterpretations \citep[e.g. a PSF artifact; a partially-subtracted disk; PAH emission, see][]{Biller2014}. 

In contrast, key features of the protoplanet candidate around HD 100546  \citep[HD 100546 b;][]{Quanz2013b}, make it more amenable to follow up and confirmation\footnote{Note that exoplanets.eu equates ``HD 100546 b" with the proposed companion at 10 AU \citep{Brittain2014}.  Throughout, we follow the naming conventions from \citet{Quanz2013b}.}.  
HD 100546 was directly imaged in a single epoch (30 May 2011) at $\sim$ 5 $\lambda$/D ($r$ $\sim$ 0\farcs{}48 or $r_{\rm proj}$ $\sim$ 47 AU), not $\sim$ 1--2 $\lambda$/D,  just exterior to a region of polarized disk emission \citep{Avenhaus2014}. 
High-resolution near-IR CO and OH spectroscopy provides evidence for another, hitherto unseen, companion at $r$ $\approx$ 10 AU \citep{Brittain2014}.  ALMA data show evidence for two rings of mm-sized dust, possibly trapped by these candidate protoplanets \citep{Walsh2014}.  At  regions exterior to the likely jovian planet formation region ($r$ $>$ 100 AU),
 the disk may also exhibit planet-induced structure \citep{Grady2005,Boccaletti2013}. 

 New observations with a different telescope/set-up can 1) confirm HD 100546 b's existence as an embedded point source-like object  and 2) identify additional signatures of protoplanets from thermal IR bright disk features at similar angular separations.   Like several stars with imaged planets (HR 8799, $\beta$ Pic, and HD 95086), HD 100546 is an intermediate-mass star and thus provides a critical look at the formation of super-jovian planets at moderate/wide separations.  

In this Letter, we report the recovery/confirmation of embedded protoplanet candidate HD 100546 b from Gemini/NICI $L^\prime$ data at the same location and brightness as reported in \citet{Quanz2013b}.    Additionally, we identify a new thermal IR-bright disk feature at roughly the same angular separation, possibly linked to another actively-forming planet.
\section{Observations and Data Reduction}
We observed HD 100546 on 31 March 2012 with the Near-Infrared Coronagraph and Imager (NICI) on Gemini-South \citep{Chun2008} in the $K_{s}$, [3.1] and $L^\prime$ filters   (0\farcs{}0179/pixel).  A future study includes contemporaneous $K_{s}$ and [3.1] imaging of HD 100546 and presents a comprehensive analysis of the disk (Honda et al. in prep.).  Our data consist of coadded 19-second exposures taken in a three-point dither pattern in \textit{angular differential imaging} mode \citep{Marois2006} (hour angle = [-3.6,-0.86]), yielding 49.5\arcdeg of parallactic motion.  The stellar PSF was stable and sky transmission was typically constant to within 5--10\%.   We identified 2128 s of good-quality science data.  
Basic image processing steps followed our previous methods for reducing thermal IR data \citep[][]{Currie2011a,Currie2014c}.  
For each image, we subtracted a moving-box median filter of width 5 $\lambda$/D to remove low spatial frequency noise and axisymmetric disk halo emission.

We performed PSF subtraction using the A-LOCI pipeline \citep{Currie2014c}, subtracting each science frame by a linear combination of \textit{reference} images obtained at different parallactic angles ($\Delta$PA $\ge$ $\delta$$\times$FWHM), where the images' coefficients are determined by solving the set of linear equations $\textbf{x}$ = \textbf{A}$^{-1}$\textbf{b}.
For  a second, independent approach we reduced the data using the Kar\'hoeven-Lo\'eve Image Projection (KLIP) algorithm \citep{Soummer2012}.   We construct eigenimages in annular regions, retaining $n_{\rm pca}$ principal components, where $n_{\rm pca}$ $<$ $n_{images}$.  As with A-LOCI, we impose a rotation gap criterion ($\delta$).

\section{Detections}

\begin{figure}
\centering
\includegraphics[scale=0.4,trim=30mm 0mm 30mm 0mm,clip]{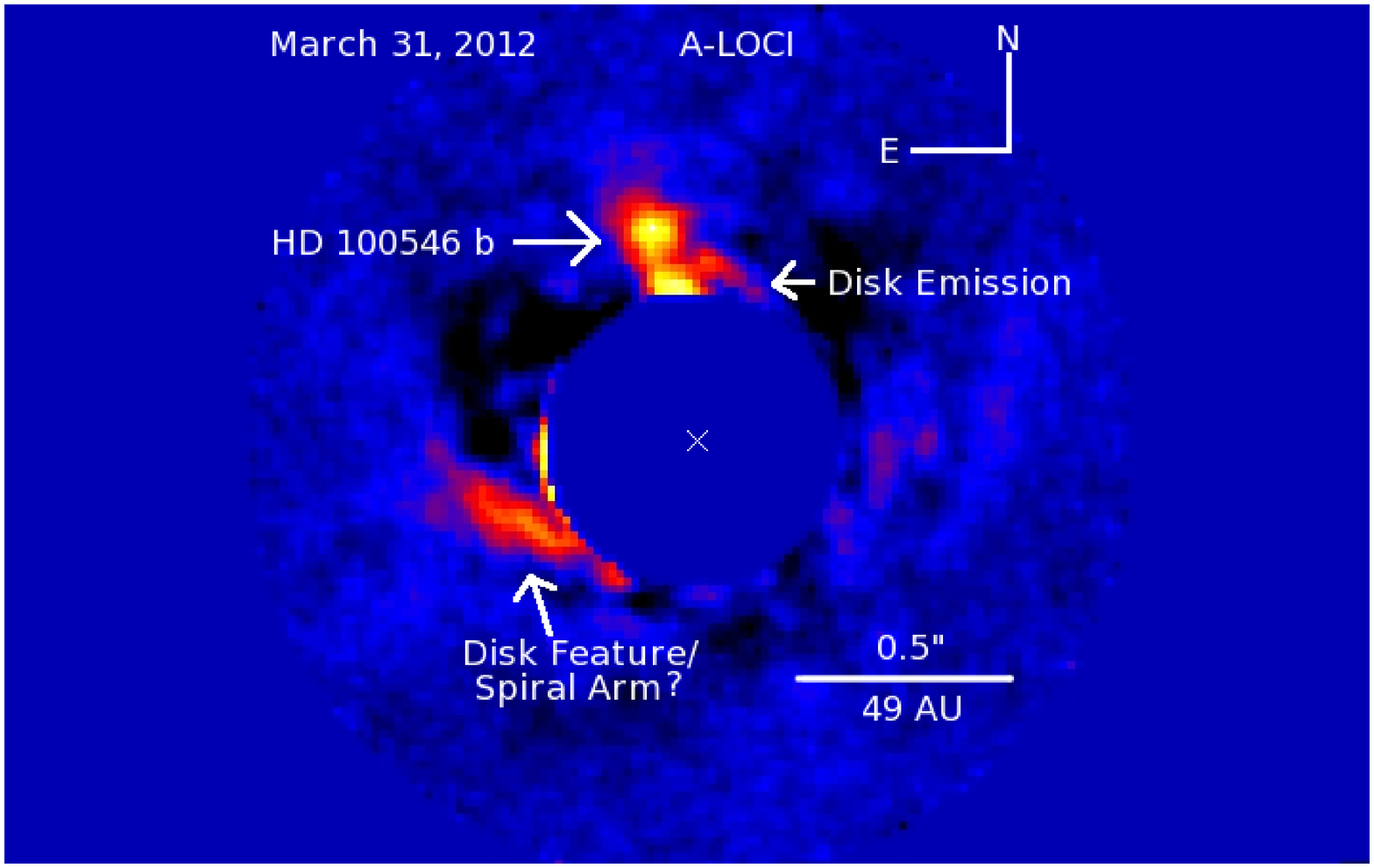}
\includegraphics[scale=0.4,trim=30mm -2mm 30mm 2mm,clip]{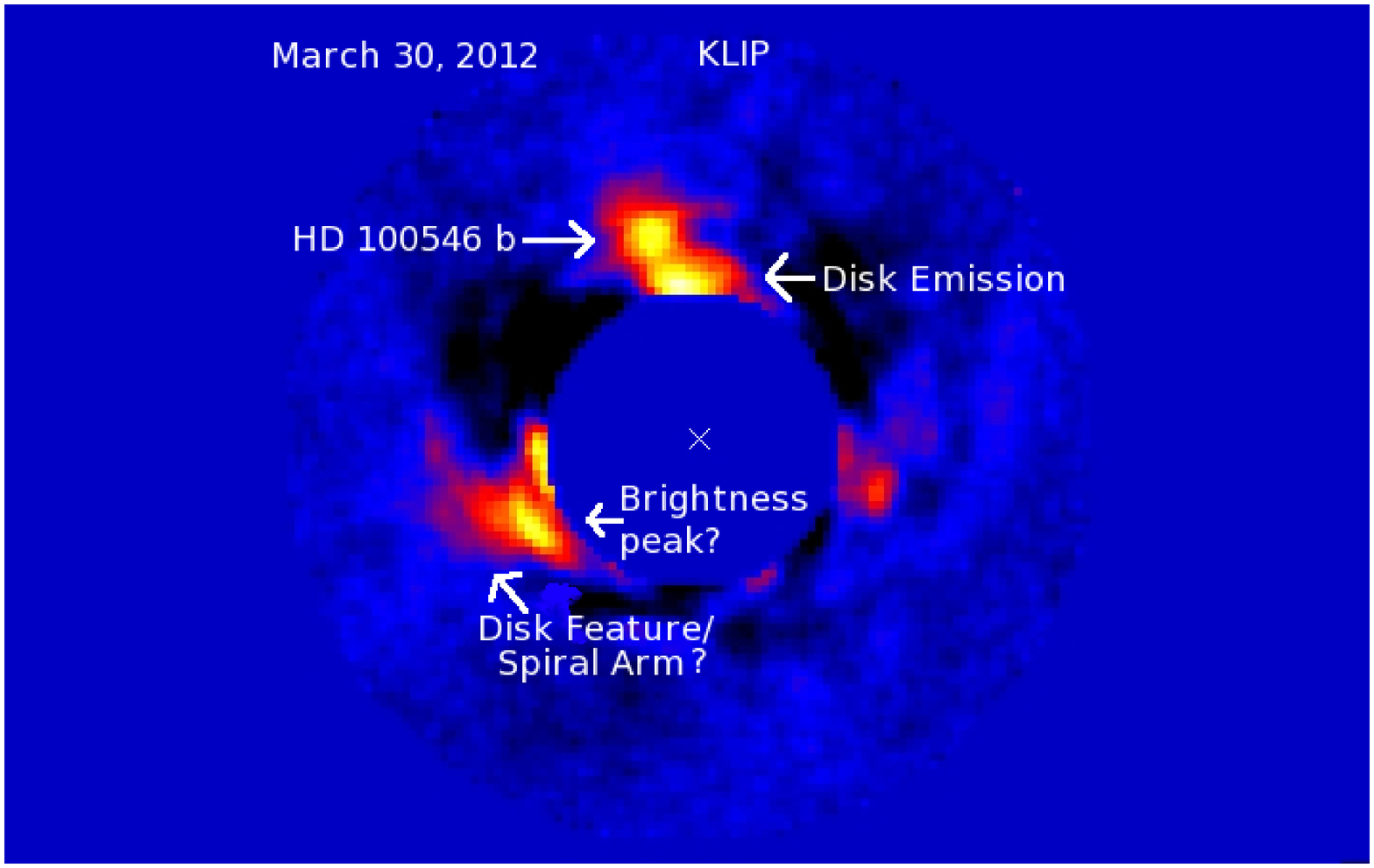}\\
\includegraphics[scale=0.4,trim=30mm 0mm 30mm 2mm,clip]{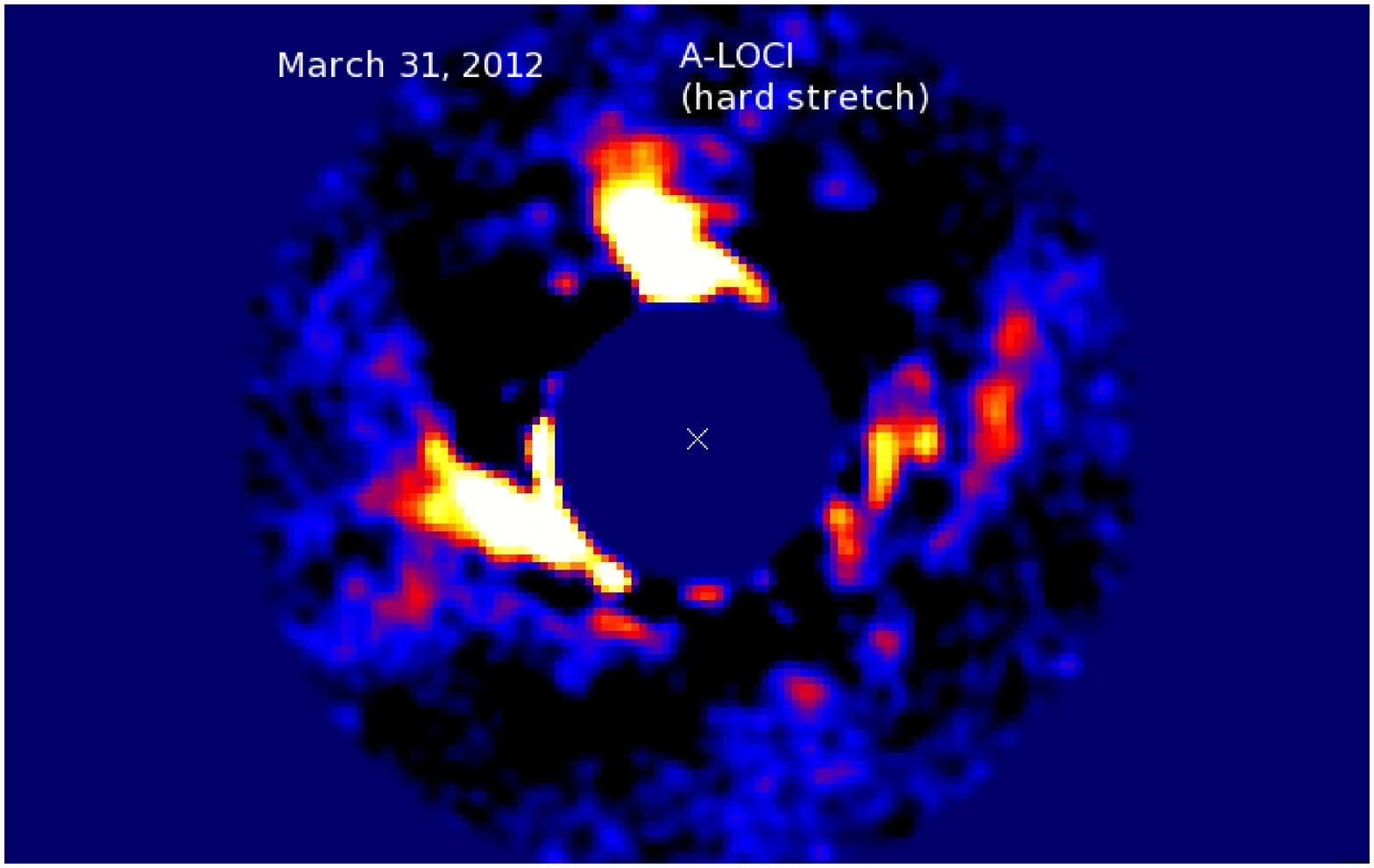}
\includegraphics[scale=0.4,trim=30mm -1mm 30mm 2mm,clip]{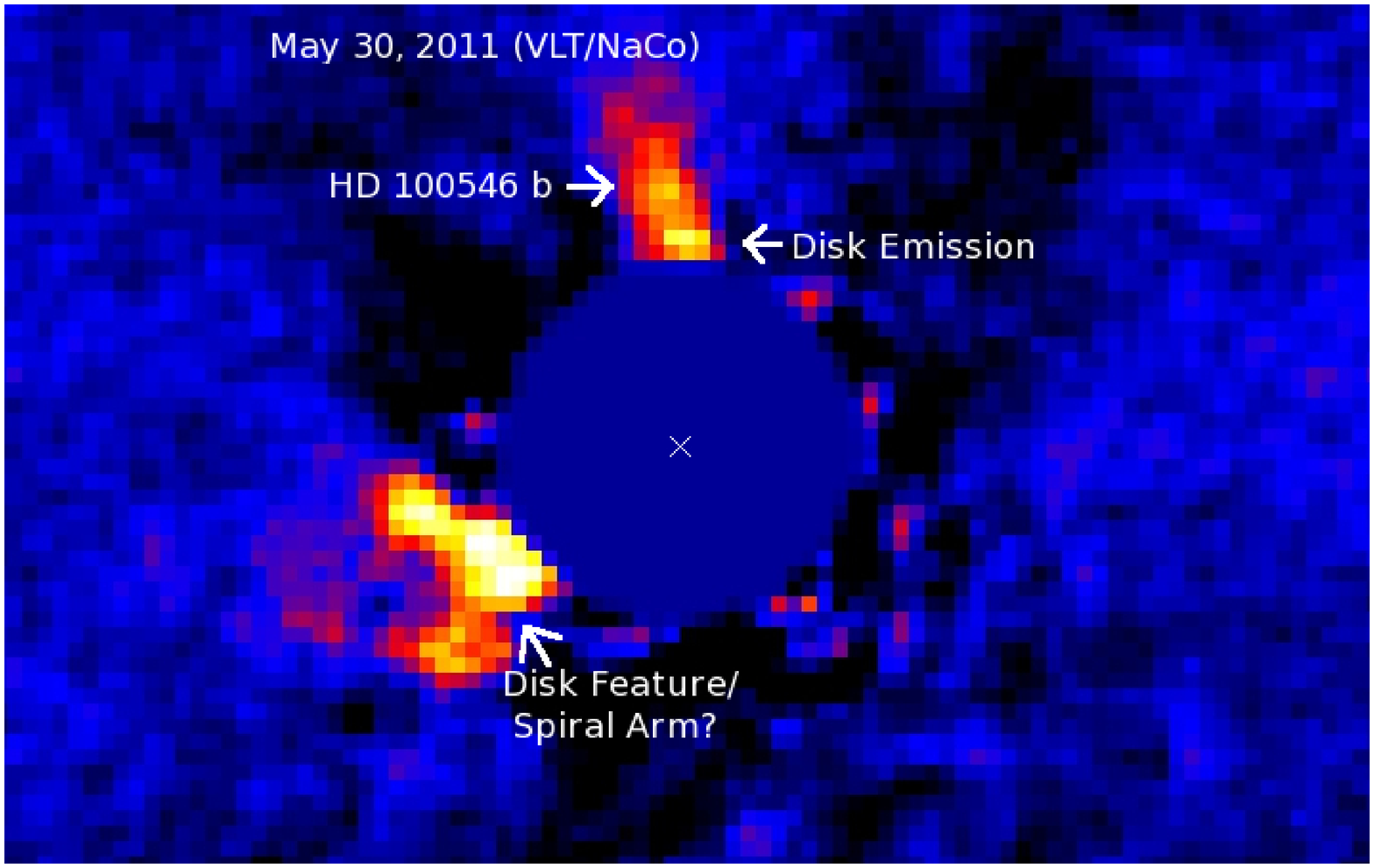}
\caption{Gemini/NICI $L^\prime$ image of HD 100546 processed with A-LOCI (top-left) and KLIP (top-right) revealing HD 100546 b and a hitherto undetected disk feature/candidate spiral arm.    The A-LOCI reduction adopts a small rotation gap ($\delta$ = 0.55) and optimization area ($N_{a}$ = 40) \citep[see ][]{Lafreniere2007} but a large SVD cutoff (7.5$\times$10$^{-4}$), aggressive frame selection ($n_{\rm ref}$ = 25), and a moving pixel mask, resulting in good throughput \citep[see][]{Marois2010b,Currie2012b,Currie2014c}.  The KLIP reduction assumes $\delta$ = 0.9, where the eigenimages are constructed over 10-pixel wide annuli at a time from the first 5 principal components \citep[see ][]{Soummer2012}.  
(bottom-left) A-LOCI image with a hard color stretch to better show the extent of the disk emission.  
(bottom-right) We identify this same extended disk features in archival 2011 NaCo $L^\prime$ data presented by \citeauthor{Quanz2013b}.}
\label{images}
\end{figure}

Figure \ref{images} displays our images reduced using A-LOCI (left) and KLIP (right), showing a point source-like peak at the position of HD 100546 b \citep{Quanz2013b} superimposed upon an extended finger of disk emission previously seen in total and polarized intensity  \citep{Quanz2013b,Avenhaus2014}.   Both reductions also reveal a bright, spiral arm-like disk structure 90 degrees away, starting at $r$ $\sim$ 0\farcs{}33, $PA$ $\approx$ 150\arcdeg  near the disk semi-major axis \citep{Avenhaus2014} to $r$ $\sim$ 0\farcs{}6, $PA$ $\approx$ 90\arcdeg.  Furthermore, the KLIP reduction (and A-LOCI reduction to a lesser extent) shows what appears to be a brightness peak superimposed on the structure at 0\farcs{}4--0\farcs{}45\footnote{
 \citet{Avenhaus2014} identify a candidate spiral at $r$ $\sim$ 0\farcs{}2--0\farcs{}3.  Our A-LOCI reduction reveals a similar feature appearing to curve in the direction of HD 100546 b, its signal is not statistically significant.  Thus, we do not consider it in this paper.}. 
 Negative subtraction footprints flank the positions of HD 100546 b and the disk feature,  consistent with them being real astrophysical features attenuated by image processing \citep[e.g.][]{Marois2010b,Milli2012,Brandt2013}.

\begin{figure}
\centering
\includegraphics[scale=0.35,trim=30mm 0mm 30mm 0mm,clip]{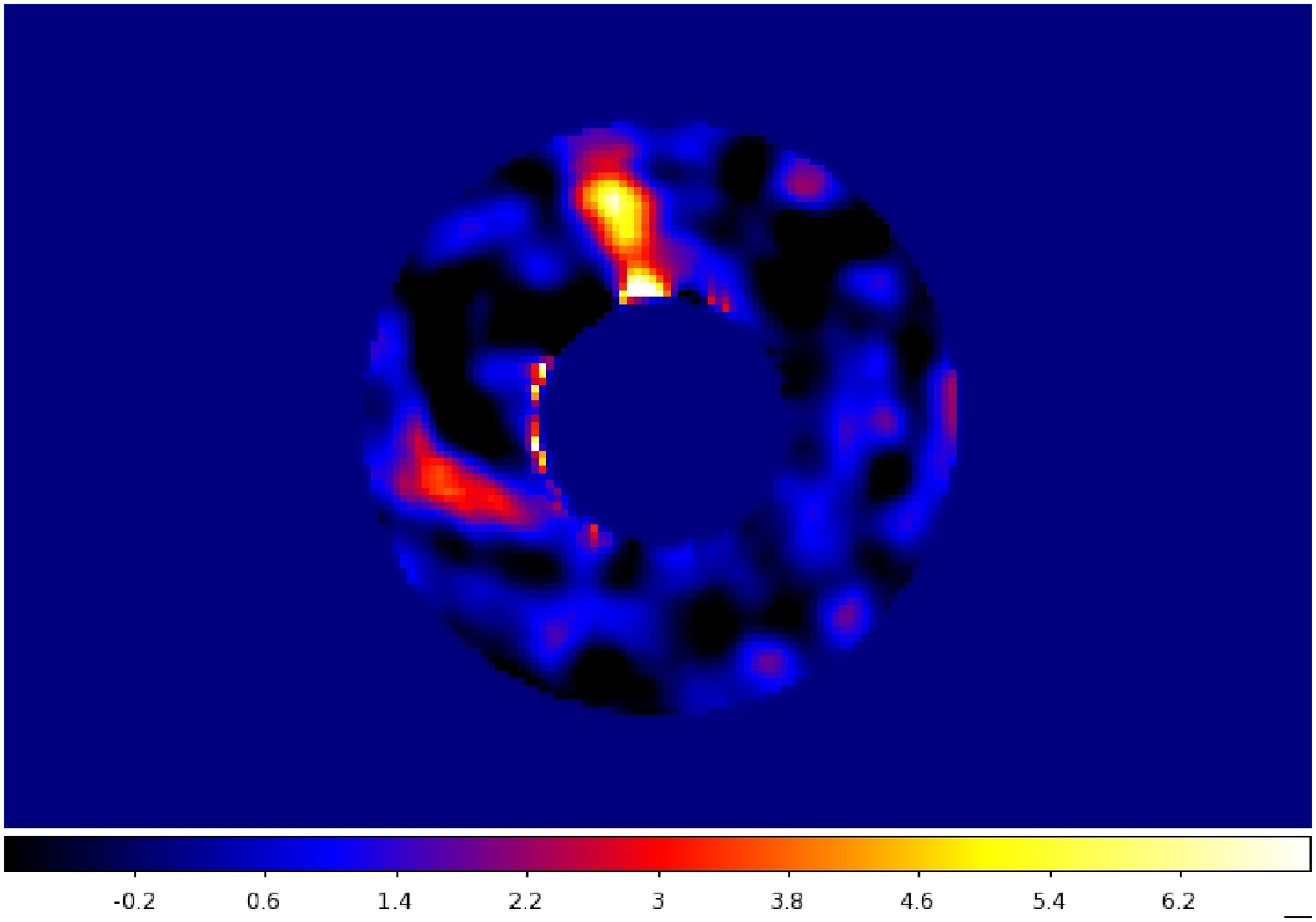}
\includegraphics[scale=0.5,trim=0mm 0mm 0mm 0mm,clip]{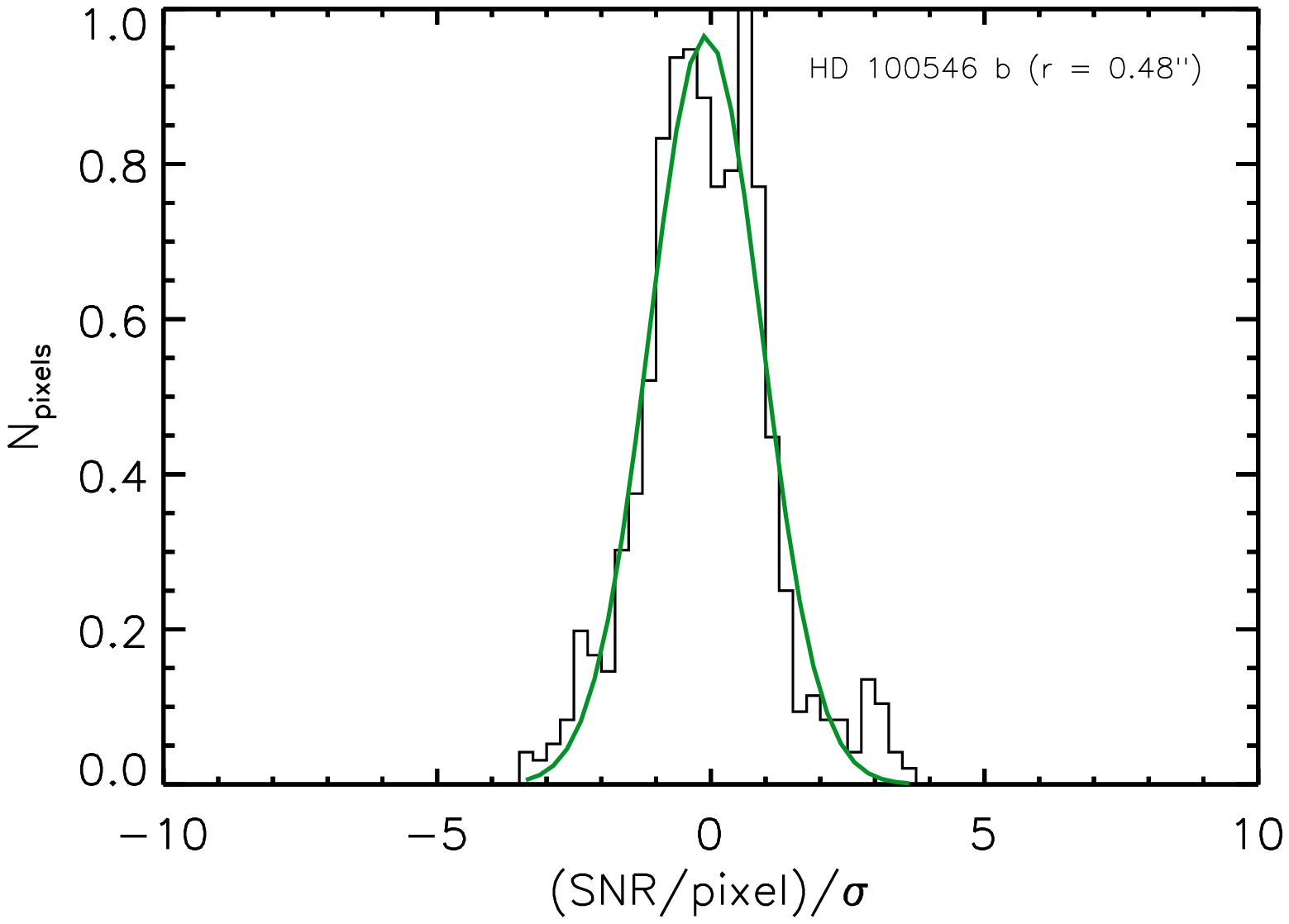}
\caption{(Left) Signal-to-noise map for the A-LOCI reduction: the map derived from our KLIP reduction is similar.  HD 100546 b (SNR = 5.3)  sits in an extended finger of emission stretching from 0\farcs{}3 to at least 0\farcs{}75.   The map also clearly reveals the disk structure/spiral arm.  (Right) Histogram distribution of pixels bracketing the separation of HD 100546 b ($r$ $\sim$ 0\farcs{}44--0\farcs{}54) with companion's signal masked.  The residuals follow a gaussian-like distribution.  Most points at 3--4$\sigma$ trace the spiral arm signal, not residual speckles.}
\label{snrtest}
\end{figure}

 To compute signal-to-noise (SNR), we use standard methods, replacing each pixel with the signal integrated over an aperture equal to that of a point source (FWHM $\sim$ 0\farcs{}1) and comparing this to the integrated signal at the same separation but other position angles \citep[e.g.][]{Thalmann2009,Currie2014c}.  Our A-LOCI and KLIP reductions recover HD 100546 b at SNR = 5.3 (Figure \ref{snrtest}, left panel) and 4.0, respectively.   
 
 In both reductions, the disk feature appears as a clear SNR $\gtrsim$ 3 trail beyond $\sim$ 0\farcs{}3.    We also detected it in archival 2011 NaCo $L^\prime$ data (Figure \ref{images}, bottom-right panel) published by \citet{Quanz2013b}.  Thus, the feature is not a PSF artifact\footnote{We also easily recover HD 100546 b and the disk feature in unpublished 2013 $L^\prime$ NaCo data  (PI. S. Quanz).  Other authors will present these results.}.  

Our SNR ratios are conservative and our detection of HD 100546 is probably comparable to the original VLT/NaCo detection from \citet{Quanz2013b}
\footnote{\citet{Quanz2013b} report a SNR $\sim$ 15 detection using a different definition for SNR, 
$S/N \sim S_{\rm integrated}/(\sigma_{\rm pixel} \times \sqrt{\pi \times r^2_{\rm aperture}}$), which is valid for noise uncorrelated
 at the pixel-to-pixel level (e.g. photon noise).  This condition is generally not satisfied in high-contrast imaging data, where both quasi-static speckles in raw images and
the power spectrum of \textit{residual} speckle noise in processed images peaks at scales much larger than one pixel \citep[Figure 9 in ][]{Brandt2013}: the relevant size scale is the resolution element, $\lambda$/D \citep{Mawet2014}.   Using \citeauthor{Quanz2013b}'s formalism, our SNR increases to SNR $\sim$ 15.2 but yields 9 other 5+$\sigma$ peaks within 0\farcs{}75.}.
We do not mask the negative subtraction footprints, nor do we mask the candidate spiral when determining the SNR for HD 100546 b and vice versa.   The radial profile of the residual noise is flat at $r$ $\sim$ 0\farcs{}4--0\farcs{}5 but drops steeply at $r$ $>$ 0\farcs{}5 (overlapping with the outer half of HD 100546 b's PSF).  Fixing the centroid position to be one pixel (two pixels) further from the star increases the SNR in the A-LOCI and KLIP reductions to 6.2 (7) and 5.0 (5.6), respectively.   


Finally, following \citet{Currie2014c}, we computed the probability distribution function (PDF) for elements on our signal-to-noise map within a 1 $\lambda$/D-wide ring enclosing HD 100546 b  (Figure \ref{snrtest}, right panel).    
The intensity residuals are almost perfectly normally distributed.   While the finite number of resolution elements at HD 100546 b's separation ($\sim$ 30) results in a somewhat higher false alarm probability than would be inferred from a perfect gaussian distribution \citep{Mawet2014}, masking real features (e.g. the disk) would again narrow the distribution, increase the SNR, and slightly lower the false alarm probability.    Using Eq. 9 in \citet{Mawet2014}, the SNR exceeds 5-$\sigma$. 



\section{Analysis}
\subsection{Properties of HD 100546 b}
\subsubsection{Size of Emitting Area and Position}
We computed  HD 100546 b's apparent FWHM from moving-box median-filtered versions of both reductions (to remove residual disk emission) following \citet{Kenyon2014}.  HD 100546 b is slightly extended compared to the predicted size of a point source ($\sim$ 0\farcs{}1 or 5.6 pixels):  FWHM$_{x,y}$ = 6.8, 7.6 pixels (0\farcs{}12, 0\farcs{}14) for the A-LOCI reduction and FWHM$_{x,y}$ = 7.2, 7.6 (0\farcs{}13, 0\farcs{}14) pixels for the KLIP reduction, implying a diameter of up to 12--13 AU.  Both processing methods can slightly attenuate emission in the azimuthal direction \citep[e.g.][]{Marois2010b}.  A lack of a good PSF model (see below) precludes our ability to better estimate HD 100546 b's size and its debias astrometry.   However, using synthetic point sources with comparable \textit{apparent} sizes, we verified that attenuation is extremely weak at HD 100546 b's separation (throughput $\approx$ 70, 85\% for A-LOCI and KLIP).  

HD 100546 b's position is [E,N]\arcsec{} =  [0.083,0.477]\arcsec{} $\pm$ 0\farcs{}012 for the A-LOCI reduction and [0.084,0.468]\arcsec{} $\pm$ 0\farcs{}013 for the KLIP reduction ($r_{\rm proj}$ $\sim$ 47 AU, $\theta$ $\sim$ 10\arcdeg):  
nearly identical to those reported by \citet{Quanz2013b}.   The small time difference between \citeauthor{Quanz2013b}'s discovery observations and our data cannot  prove common proper motion.  However, HD 100546 b's position within the disk makes the companion extremely unlikely to be a background object.
\subsubsection{Brightness and Estimated Mass}


We can obtain only rough estimates for HD 100546 b's brightness.  Our photometric calibrator BS4638 (m$_{L^\prime}$ = 4.5) was observed with the AO loop off, HD 100546A's PSF core was saturated, and we were unable to obtain other unsaturated NICI PSF images.  For a model HD 100546 b PSF, we simply
 constructed a gaussian with FWHM$_{x,y}$ = [7, 7.6] pixels and 
compared the encircled energy between a 3-pixel radius aperture and an infinite one to estimate an aperture correction.    We then injected negative copies of the model PSF at the HD 100546 b's position and identified the range of magnitudes where the signal is clearly oversubtracted/undersubtracted, defining the estimated magnitude as the range's midpoint and equating the intrinsic photometric uncertainty with this range.   
 Photometric errors consider HD 100546 b's SNR, the photometric calibrator's SNR,  and uncertainties in the throughput correction.
We derive m$_{L^\prime, b}$ = 13.06 $\pm$ 0.51, consistent with \citet{Quanz2013b}.  

HD 100546 b plausibly has a mass well below the deuterium-burning limit.
Given an age of 5--10 $Myr$, a distance of 97 $pc$ \citep{vanLeeuwen2007}, and using standard hot-start models to estimate planet mass \citep{Spiegel2012}, our photometry and the \citeauthor{Boccaletti2013} upper limits imply a mass of $\approx$ 15 $M_{J}$.   However, HD 100546 b could be newly born, significantly younger than the star \citep[see][]{Currie2013a}.  For $t$ = 1 $Myr$, the \citeauthor{Spiegel2012} models nominally yield $M_{\rm b}$ $\sim$ 7 $M_{J}$, $<$ 10 $M_{J}$ for an initial entropy of $S_{init}$ $>$ 11.75 $k_{B}$/baryon. 

  If HD 100546 b is resolved, much (all?) of its emission could come from an accreting circumplanetary disk \citep[e.g.][]{Fujii2014,Zhu2014} and its mass could be significantly below 10 $M_{J}$.    Assuming a mass of 1 $M_{J}$, a truncated disk with $R_{\rm in}$ = 1.5--2 $R_{J}$, and a circumplanetary accretion rate to 5--10\% that of the circumstellar disk \citep[or $\dot{M_{b}}$ $\approx$ 3--6 $\times$ 10$^{-6}$ $M_{\rm J}$ $yr^{-1}$, c.f.][]{Pogodin2012}, the disk can reproduce our $L^\prime$ photometry without making HD 100546 b detectable at $K_{s}$.

\subsection{Spiral Arm Fitting of the Newly-Discovered Thermal IR-Bright Disk Structure: Evidence for a Second Wide-Separation Planet?}
Although the morphology of the newly-identified disk feature at $r$ $>$ 0\farcs{}35 lacks conclusive results, it resembles spiral arms in HD 100546's disk at other separations \citep[][]{Grady2005,Ardila2007,Avenhaus2014} and in other protoplanetary disks \citep{Muto2012,Grady2013}.   Unlike these spirals, this feature appears to be particularly bright in the thermal IR but less easily identifiable in polarized light \citep[c.f.][]{Avenhaus2014}.    Thus, any spiral features primarily trace the disk structure in total intensity, not polarized intensity.  

We model this feature using spiral density wave theory following the methodology outlined in \citet{Muto2012}, fitting the spiral shape, which depends on the 
 wave launching point ($r_{o}$, $\theta_{o}$); the disk aspect ratio ($h_{c}$) at the launching point; the radial power-law index for the rotation angular velocity, $\alpha$; and the radial power-law index for the sound speed, $\beta$.  We set $\alpha$ = -1.5 (the Keplerian value) and perform a four-parameter fit varying $r_{o}$, $\theta_{o}$, h$_{c}$, and $\beta$\footnote{Our modeling results derive just from the peak signal along the arm.  Azimuthal self-subtraction does, however, prevent us from precisely measuring the spiral amplitude, which is connected to planet mass if the spiral is launched by a planet.}.  
 
Although the disk morphologies in the A-LOCI and KLIP reductions are similar, we focus on modeling the KLIP reduction \citep[see discussion in][]{Soummer2012} and use the A-LOCI simply as a comparison.  We adopt a major axis's position angle and inclination of 138\arcdeg and 50\arcdeg, respectively \citep[][]{Avenhaus2014}.
The range of ``good-fitting solutions" result  from a $\Delta$$\chi^{2}$ criterion and correspond 
to the 5-$\sigma$ confidence level if the errors associated with representative spiral shape positions are independent.

\begin{figure}
\centering
\includegraphics[scale=0.75,trim=20mm 0mm 33mm 0mm,clip]{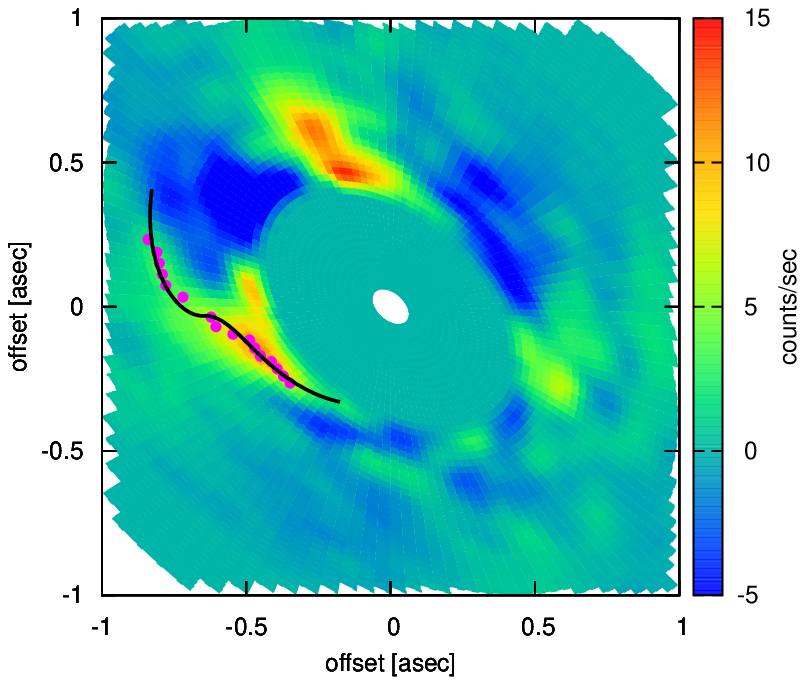}
\includegraphics[scale=0.75,trim=28mm 0mm 20mm 0mm,clip]{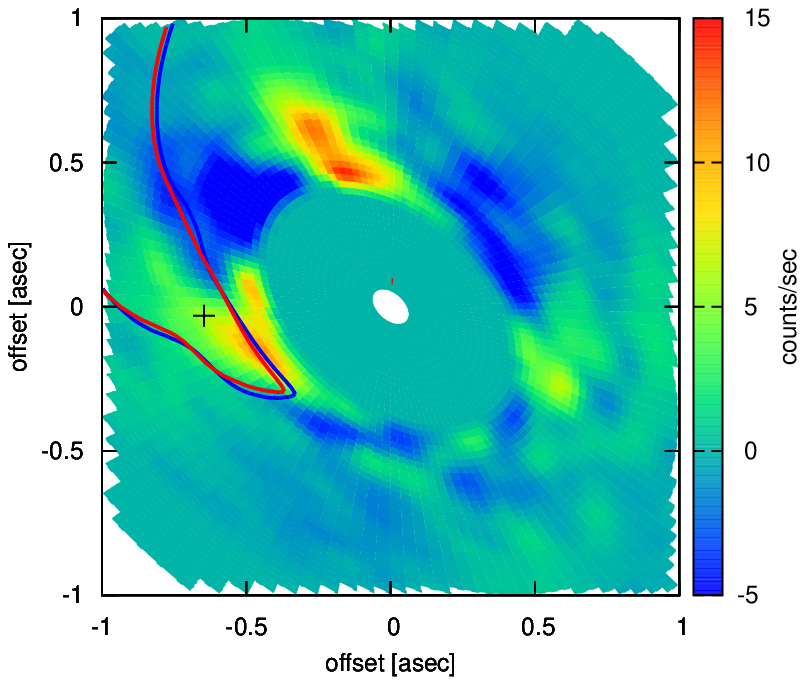}
\caption{Spiral density wave modeling.  (Left) The local maxima (magenta dots) tracing the location of the spiral from $r$ $\sim$ 0\farcs{}35 to 0\farcs{}1 (black line) on the deprojected KLIP image.  (Right) The best-fit wave launching point (cross) along with the 5-$\sigma$ contours  from the KLIP reduction (blue) and A-LOCI reduction (red).}
\label{spiralimage}
\end{figure}

Table \ref{tab_spiral} summarizes our results, while Figure \ref{spiralimage} compares our best-fit parameters to the de-projected KLIP-reduced image.   Significant model fitting degeneracies allow only modest constraints on most spiral properties.
However, the best-fit wave launching point (and the possible location of a perturbing planet) is at $r_{o}$, $\theta_{o}$ = 0\farcs{}65, 93\arcdeg, within the wave itself.   The 5 $\sigma$ contours imply that HD 100546 b is not launching this spiral.  Instead, much of the spiral arm region exterior to 0\farcs{}45 could be the site of the wave launching point, including the local maximum most clearly seen in the KLIP reduction.
Our results from modeling the A-LOCI reduced image likewise imply a wave launching point exterior to 0\farcs{}45, plausibly within the wave but not consistent with HD 100546 b's position.
Rerunning our fits using a sparse sampling where the positional errors should be (more) independent yields nearly identical results. 
If the newly-identified disk feature is a spiral arm induced by a hitherto unseen planet, then the planet is most likely an additional companion exterior to $r$ $\approx$ 0\farcs{}45.

\section{Summary and Discussion}
Our study reports the first independent, second-epoch (re-)detection of a directly-imaged protoplanet.  Using Gemini/NICI thermal IR data, we recover  HD 100546 b at the same location and brightness as in the discovery image \citep{Quanz2013b}.
 Furthermore, we discover an extended disk feature 90\arcdeg from HD 100546 b, resembling a spiral density wave, with a possible thermal IR peak at 0\farcs{}4--0\farcs{}45.   While interpreting this feature is not straightforward, modeling it as a spiral density wave implies a wave launching point, the possible location of another protoplanet,  exterior to 0\farcs{}45 and possibly within the visible wave itself.

Although HD 100546 b is now recovered, its status as a directly-imaged planet is on less firm ground than companions like HR 8799 bcde, since it lacks common proper motion confirmation, may be spatially extended, and could in principle be a partially-subtracted piece of the disk.  However, it is too far from the star to be reprocessed disk emission.  Its non-detection in polarized light and in near-IR total intensity disfavors explaining it as light scattered by small dust grains: if it is indeed a partially-subtracted disk feature, it is a feature with different dust properties.   PAH emission in HD 100546's disk appears to originate at the inner disk rim (13 AU), not out at r$_{\rm proj}$ $\sim$ 47 AU \citep{Geers2007}; thermal instabilities likewise seem implausible \citep[c.f.][]{Biller2014}.   

Furthermore, HD 100546 b and the polarized disk material on which it sits together cannot identify another spiral arm, since it would wind in the \textit{opposite} direction from bona fide arms at wide separations \citep{Grady2005} and our candidate arm.   After ruling out other alternatives, we conclude HD 100546 b identifies locally-produced thermal emission from either a young protoplanet or dissipation driven from some other, unknown mechanism.   HD 100546 b as a protoplanet naturally explains the ring of mm-sized dust located exterior to it \citep{Walsh2014}.   
 
HD 100546 b's nature may soon be decisively determined.   Astrometry from 2013--2014 may demonstrate that HD 100546 b is not a background object.
 New, multi-wavelength photometry (e.g. 2--5 $\mu$m) can clarify whether HD 100546 b has planet-like colors.  Integral field spectroscopy with the \textit{Gemini Planet Imager} or \textit{SPHERE} \citep{Macintosh2008,Beuzit2008} can show whether HD 100546 b has a spectrum similar to that of the youngest planetary objects \citep[e.g.][]{Bonnefoy2014, Currie2014a}.  


The newly-detected, spatially extended inner disk feature/candidate spiral arm requires new data to interpret cleanly.  Furthermore, while spiral density waves in disks are often associated with planets, a spiral could have non-planet origins (Lyra et al. 2014, A\&A submitted).   New IR observations may 
allow us to assess whether a planet-driven spiral is reasonable.  Further analysis of other regions of the HD 100546 disk from IR data (Honda et al. in prep.) and the (sub-)millimeter with ALMA  \citep[e.g.][]{Walsh2014} will better place our newly-discovered structure within the context of HD 100546's disk morphology in general, provide clues needed to better interpret HD 100546 b and the disk structure we first identify.  

Our study further cements HD 100546 as a superb laboratory for studying the formation of multiple (super-)jovian planets \citep[e.g][]{Quillen2006}.  The system contains one bona fide protoplanet candidate at $r_{\rm proj}$ = 47 AU, strong evidence for a candidate at 10 AU \citep{Brittain2014}, and potential evidence for additional planets exterior to $r_{\rm proj}$ $\approx$ 45 AU (this work).      The full range in (projected) separation is qualitatively similar to the projected separations of HR 8799 bcde \citep{Marois2010a}, the only known imaged multi-planet system.   Thus, HD 100546 provides important insight into the formation and early evolution of super-jovian planets at moderate/wide separations around intermediate-mass stars.

\acknowledgements 
We  thank Wladimir Lyra, Scott Kenyon, Mickael Bonnefoy, Christian Thalmann, Nienke van der Marel, and the anonymous referee for helpful comments.    
This research makes use of the ESO archive.

{}

\begin{deluxetable}{llllllllllllll}
 \tiny
\setlength{\tabcolsep}{0pt}
\tabletypesize{\scriptsize}
\tablecolumns{10}
\tablecaption{Spiral Arm Fitting Results}
\tablehead{{Reduction}&{$r_{o}$ (best, range)}&{$\theta_{o}$ (best, range)}&{$h_{c}$ (best, range)}&{$\beta$ (best, range)}\\
{}&{(\arcsec{})}&{(\arcsec{})}&{(${\arcdeg}$)}}
\startdata
\textit{primary}\\
KLIP & 0.65, $>$ 0.45& 93, $<$ 133& 0.2, $>$ 0.1 & 0.05, 0--1\\
\textit{comparison}\\
A-LOCI &  0.94, $>$ 0.46 & 76, $<$ 133 & 0.3, $>$ 0.1 & 1, 0--1\\
\enddata
\tablecomments{The columns list the best-fit and range for position of the co-rotation radius ($r_{o}$, $\theta_{o}$), the scale height ($h_{c}$), and the gas sound speed power law ($\beta$).} 
\label{tab_spiral}
\end{deluxetable}

\end{document}